\documentclass[aps,jmp,amsmath,amssymb,reprint]{revtex4-1}
\usepackage{float}
\usepackage{graphicx}
\usepackage{dcolumn}
\usepackage{bm}
\usepackage{graphicx}
\usepackage{subfigure}
\usepackage{physics}
\usepackage[english]{babel}
\usepackage[usenames, dvipsnames]{color}
\usepackage{hyperref}
\usepackage[normalem]{ulem}
\usepackage[utf8]{inputenc}
\usepackage{placeins}

\begin{document}

\title{Comprehensive quantum transport analysis of M-Superlattice structures for barrier infrared detectors}          

	\author{Anuja Singh}
	
		\author{Swarnadip Mukherjee}

	\author{Bhaskaran Muralidharan}
	\affiliation{Department of Electrical Engineering, Indian Institute of Technology Bombay, Powai, Mumbai-400076, India}
\date{\today}

\begin{abstract}
In pursuit of designing superior type-II superlattice barrier infrared detectors, this study encompasses an exhaustive analysis of utilizing M-structured superlattices for both the absorber and barrier layers through proper band engineering and discusses its potential benefits over other candidates. The electronic band properties of ideally infinite M-structures are calculated using the eight band $k.p$ method which takes into account the effects of both strain and microscopic interface asymmetry to primarily estimate the bandgap and density-of-states effective mass and their variation with respect to the thicknesses of the constituent material layers. In contrast, for practical finite-period structures, the local density-of-states and spectral tunneling transmission and current calculated using the Keldysh non-equilibrium Green's function approach with the inclusion of non-coherent scattering processes offer deep insights into the qualitative aspects of miniband and localization engineering via structural variation. Our key results demonstrate how to achieve a wide infrared spectral range, reduce tunneling dark currents, induce strong interband wavefunction overlaps at the interfaces for adequate absorption, and excellent band-tunability to facilitate unipolar or bipolar current blocking barriers. This study, therefore, perfectly exemplifies the utilization of 6.1{\AA} material library to its full potential through the demonstration of band engineering in M-structured superlattices and sets up the right platform to possibly replace other complex superlattice systems for targeted applications.
\end{abstract}


\maketitle

\section{INTRODUCTION}
The evolution of materials technology in the field of infrared (IR) detection has experienced a rapid turnaround in the late nineties when traditional bulk materials such as HgCdTe, InSb were replaced by quantum-engineered structures as the former exhibited limited applicability due to the non-uniform growth defects and their ineffectiveness to cover the full IR range \cite{plis2014inas,rogalski2018antimonide,martyniuk2014new,martyniuk2014barrier}. Since then, type-II superlattices (T2SL) composed of InAs/(In,Ga)Sb heterostructures have gained a lot of attention for their thickness dependent bandgap tunabilities, suppressed Auger recombination and high quantum efficiencies. This has enabled them to become the latest industry-standard technology for next-generation IR detectors and focal plane arrays (FPA) \cite{plis2014inas,mukherjee2021carrier,rogalski2018antimonide,ting2020long,zavala2020antimonide,martyniuk2014new}. Subsequently, one of the major breakthroughs en-route achieving a higher operating temperature and background limited IR detection has come in effect in the mid-2000s, when the concept of barrier-based detectors was first introduced. This later laid out an ideal platform to explore several novel device architectures of the form nBn, pBp, pBiBn etc., through ``band-diagram engineering" \cite{rogalski2017inas,nguyen2007type,nguyen2008band,plis2014inas,zavala2020antimonide}. \\
\indent Despite numerous advantages, T2SL systems suffer from an inadequacy to produce perfect hole barriers due to the strong adherence of hole minibands to the bottom of GaSb valence bands and the lack of tunability with varying GaSb layer thicknesses \cite{razeghi2010band,nguyen2007dark,nguyen2009minority,zavala2020antimonide,lang2013electronic}. Moreover, the spatial separation of carrier localizations in these structures leads to smaller absorption, affecting the photo response \cite{ting2020long}. To overcome this, complex structures like M, W, N were introduced later by inserting an additional high-bandgap AlSb layer in the original T2SL layout, which offered higher degrees of freedom to achieve a full tailorability of miniband alignment and better wavefunction engineering \cite{ting2020long,lang2013electronic,nguyen2008band,maimon2006n}.\\
\indent The placement of the AlSb layer determines the nomenclature of these structures based on the shape formed on joining the conduction band edges \cite{ting2020long,nguyen2008band}. In M-superlattice (MSL) structures, the AlSb layer is inserted at the center of the GaSb layer which splits the single hole quantum well into two and makes the valence band more sensitive to the GaSb width \cite{lang2013electronic,ting2020long,nguyen2007type,nguyen2007dark}. Furthermore, due to the AlSb layer, the center of the hole wavefucntion in the GaSb layer shifts closer to the center of electronic wavefunction in the InAs layer which effectively increases the overlap between them, thereby facilitating higher absorption \cite{ting2020long,lang2013electronic}. In addition, MSLs possess fine tuning of band alignments between the absorber and the barrier and strongly provide an impediment to tunneling transport of carriers, thereby knocking off their contribution to the SRH dark current \cite{nguyen2007dark,nguyen2007type,nguyen2008band,nguyen2009minority}. However, a unified study on the key roles played by the AlSb layer width in modulating the bandgap, density of states (DOS) effective masses, valence miniband offsets, interband overlaps, and spectral transport are still scarce in the literature and thus form the basis of this study.\\
\indent This motivates us to perform a scrupulous theoretical investigation of MSL systems and present a thorough comparison with the conventional T2SL systems \cite{li2010intrinsic,livneh2012k}. We show that the bandgap of such systems can be accurately predicted by incorporating an appropriate interface model that takes into account the effects of strain due to lattice mismatch and microscopic interface asymmetry (MIA) corresponding to the alternate arrangement of interface materials \cite{li2010intrinsic,szmulowicz1996numerically,szmulowicz2004effect,hong2009applicability,krebs1996giant,mlinar2005nonsymmetrized,szmulowicz2004effect}. In addition, we emphasize the non-trivial roles played by the inserted AlSb layer in tandem with the original T2SL material layers in regulating the miniband edges and carrier effective masses towards an efficient engineering of carrier transport to optimize the dark current. In subsequent explorations, we compare finite-period MSL and T2SL systems in terms of the tunneling transmission and local density of states (LDOS) features evaluated from the Keldysh non-equilibrium Green's function (NEGF) approach and highlight key differences based on the nature of carrier localization in the available subband states \cite{priyadarshi2018superlattice,mukherjee2018improved,mukherjee2021carrier,tibaldi2021modeling,bertazzi2020nonequilibrium}. The strong interband overlap around the interface region and a signature of spatially continuous broad current spectra in MSL structures are respectively indicative of enhanced absorption and phonon-assisted non-coherent miniband transport of carriers. Finally, MSL structures are shown to demonstrate excellent and precise band-tunability features through carefully introduced design guidelines which makes them suitable to be used as carrier blocking unipolar or bipolar barriers.\\
\indent The rest of the paper is organized as follows. In Sec. \ref{sec_theory}, we describe the $\bf{k.p}$ and NEGF theory applied for modeling electronic band properties and miniband characteristics of MSLs. We discuss the results in Sec. \ref{sec_result} which is divided into three subsections. Section \ref{eg_kp} discusses the band structure properties evaluated via the  $\bf{k.p}$ method and explains the effect of varying layer widths on the key band parameters. In Sec. \ref{comp}, a comparative study between a finite MSL and a finite T2SL absorber is presented in terms of the local density of states (LDOS) and spectral tunneling transmission evaluated via the NEGF approach. Finally, Sec. \ref{Barrier} depicts the potential of MSL structures to be used as both electron and hole barriers with respect to the T2SL based absorbers. The paper is concluded in Sec. \ref{conclu}.
\section{THEORETICAL APPROACH}
\label{sec_theory}
\subsection{ $\bf{k.p}$ method}
\label{sec_1}
The electronic band structure of the non-common atom (NCA) interface InAs/GaSb based T2SL is commonly calculated by the k.p based envelope function approximation (EFA) approach \cite{szmulowicz1996numerically,szmulowicz1998numerically}. To account for the strain and interface effects, an additional layer of InSb is inserted at the interface of InAs and GaSb \cite{delmas2019comprehensive,mukherjee2021carrier,livneh2012k}.
\begin{figure}[!htbp]
	\centering
	{\includegraphics[height=0.2\textwidth,width=0.4\textwidth]{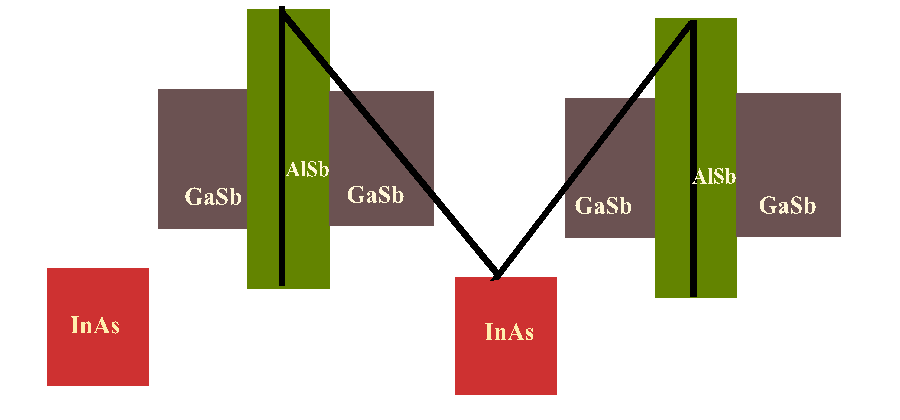}\label{M Structur}}
	\caption{Energy band alignment of the 6.1{\AA} family of compound semiconductors (InAs, GaSb and AlSb): The given arrangement of materials as shown here, forms a M-shaped structure within a typical unit cell of the periodic SL. In the absence of the AlSb layers, the structure becomes a conventional InAs/GaSb T2SL.
	}
	\label{Eb}
	\end{figure}
However, the EFA model is incapable of distinguishing the CA and NCA type interfaces leading to an overestimation of the bandgap \cite{li2010intrinsic}.
Therefore, to investigate these NCA superlattices, we include the model proposed by Krebs and Voisin with the EFA model, which distinguishes the interface chemical bonds stacked in forward and backward directions \cite{krebs1996giant,li2010intrinsic}. In order to take the MIA effect into consideration
within the $\bf{k.p}$ framework, the two distinct interfaces, GaSb-on-InAs and InAs-on-GaSb are considered \cite{li2010intrinsic,hong2009applicability,krebs1996giant,mlinar2005nonsymmetrized,szmulowicz2004effect}. The explicit Hamiltonians for both interface and strain are added with the  original $\bf{k.p}$ Hamiltonian. The total Hamiltonian  of the system can thus be written as $H(k)=H_{k}+H_{S}+H_{IF}+H_{strain}$, where, $k$ is the in-plane wave vector, $H_{k}$ are $H_S$ are respectively the $\bf{k}$ and spin dependent terms, $H_{strain}$ stands for the substrate induced strain effects due to the lattice mismatch \cite{li2010intrinsic,livneh2012k,delmas2019comprehensive,becer2019modeling}, and $H_{IF}$ is the interface term which accounts for the NCA interface MIA effects, given by
\begin{equation}
	H_{4}^{I}=H_{XY}^{AB/BA}\begin{bmatrix}
		0 & 0 & 0 & 0  \\ 0 & 1 & 1 & 0  \\
		0 & 1 & 1 & 0 \\ 	0 & 0 & 0 & 1
	\end{bmatrix},
\end{equation}
\begin{equation}
	H_{IF}=\begin{bmatrix}
	H_{4}^{I}	 & 0   \\ 0 & H_{4}^{I}
	\end{bmatrix}.
\end{equation}
Here, $H_{IF}$ is added at each interface by the delta interface strength potentials i.e $H_{XY}^{BA}=490meV$ and  $H_{XY}^{AB}=870meV$, respectively denoting the InSb (GaSb-on-InAs) and GaAs (InAs-on-GaSb) like strength potentials at the interface \cite{li2010intrinsic,livneh2012k,li2010intrinsic,lang2011interface,le2019simulation}.
For MSL, the GaSb layer is treated as unstrained and the other layers i.e InAs and AlSb as strained to attain the GaSb lattice constant \cite{lang2011interface}. The strain Hamiltonian contains the Pikus-Bir deformation potential (b, ac, av) \cite{livneh2012k,lang2011interface} and the strain terms, given by
\begin{equation}
\centering
\epsilon_{xx}=\epsilon_{yy}=\frac{a_{GaSb}-a}{a},
\end{equation}
\begin{equation}
\centering
\epsilon_{zz}=-2\frac{C_{11}}{C_{12}}\epsilon_{xx},
\end{equation}
where, $a_{GaSb}$ is the lattice constant of GaSb and $a$ is the lattice constant of the layer for which strain parameters are to be calculated, $\epsilon_{xx} $, $\epsilon_{yy} $, and $\epsilon_{zz} $ are strains along x, y and z direction, and $C_{11}$, $C_{12}$ are the elastic stiffness coefficients \cite{livneh2012k}. The interband momentum matrix Kane's parameter, being ideally taken same throughout the lattice, is assumed to be an weighted average of Kane's energy of InAs, GaSb and AlSb with respect to their thicknesses \cite{lang2011interface}. Also, a finite difference discretization technique is employed with the periodic boundary conditions to solve the slow varying envelope functions \cite{mukherjee2021carrier,galeriu2005k}. The schematic of the MSL structure as shown in Fig. \ref{Eb}, contains the additional AlSb layers at the center of each GaSb layer throughout the lattice. The interface matrix ($H_{IF}$) is only considered at the NCA interfaces, and the other material parameters for InAs, GaSb, and AlSb  taken for the simulation of MSL electronic band structure are provided in Table \ref{table1}.
\begin{table}
	\centering
	\caption{ Material parameters of InAs, GaSb and AlSb used for the $\bf{k.p}$ electronic band structure calculation at a temperature of $77K$ \cite{becer2019modeling,livneh2012k,delmas2019comprehensive,qiao2012electronic,mukherjee2021carrier,vurgaftman2001band}}
 	\begin{tabular}{|c|c|c|c|}
		\hline
		Parameters & InAs & GaSb & AlSb \\
		\hline
		Lattice constant({\AA})& 6.0584& 6.0959 & 6.1297\\
		\hline
		Energy band gap at 0K ($eV$)& 0.418 & 0.814 & 2.386\\
		\hline
		Elastic stiffness constant ($C_{11}$) &  8.329 & 8.842 & 8.769\\
		($10^{11} dyne/c{m^2}$) & & &\\
		\hline
		Elastic stiffness constant ($C_{12}$) &  4.526 & 4.026 & 4.341\\
		($10^{11} dyne/c{m^2}$) & & & \\
		\hline
		Deformation potential $ac$ ($eV$) &  -5.08 & -7.5 & -4.5\\
		\hline
		Deformation potential $av$ ($eV$) & 1 & 0.8 & 1.4\\
		\hline
		Deformation potential $b$ ($eV$) & -1.8 & -2 & -1.35\\
		\hline
		Varshini Parameter $\alpha $ $[meV/K$]& 0.276 & 0.417&0.42\\
		\hline
		Varshini Parameter $\beta$ [$K$]&93&140&140\\
		\hline
		Effective mass electron ($m_e^*$)& 0.022&0.0412& 0.14\\
		\hline
		Luttinger parameter $\gamma1$  & 19.4& 11.84 & 4.15\\
		\hline
		Luttinger parameter $\gamma2$  & 8.545 & 4.25\ & 1.28\\
		\hline
		Luttinger parameter $\gamma3$  & 9.17 & 5.01 & 1.75\\
		\hline
		Interband mixing parameter Ep [$eV$] & 21.5 & 22.4 & 18.7\\
		\hline
		Spin orbit splitting (SO) [$eV$] & 0.38& 0.76 & 0.65\\
		\hline
		Valence band offset (VBO) [$eV$] & -0.56 & 0 & -0.38\\
		\hline
	\end{tabular}
	\label{table1}
\end{table}
\subsection{Keldysh NEGF method}
\label{sec_2}
In quantum-confined finite SL structures, the LDOS and electronic transmission function offer a concrete and qualitative understanding on the quantum mechanical nature of the carrier localization profile \cite{aeberhard2018photocarrier,miniband_spie} and spectral current flow under different biasing conditions. In this work, we employ the quantum transport based NEGF formalism \cite{dattaLNE,DattaQT} as a mathematical tool to numerically compute these parameters, incorporating the concerned device physics models \cite{Henrickson,Aeberhard_jce,QCLnegf,AeberhardPRB2008,PRL_nanotube,myTED,myPRA,mukherjee2018improved,priyadarshi2018superlattice,mukherjee2021carrier}. The retarded Green's function ($G$) with the proper information of self energies corresponding to the macroscopic contacts and scattering phenomena along the longitudinal energy (E) is defined as \cite{Cavassilas_NEGF,akhavan_effectivemass,akhavanted,Kolek_effmass,foreman1993effective},
\begin{equation}
G(z,z',E)=[E^+I-H-U-\sum_{j}\Sigma^C_{j}(E)-\Sigma^S(E)]^{-1},
\label{green}
\end{equation}
where, $E^+=E+i\eta^+$ where $\eta^+$ is a small positive number, $z,z'$ are the position indices, $I$ denotes the Identity matrix, $H$ represents the 1-D tight-binding Hamiltonian matrix of the SL in real space and $U$ is the potential profile calculated from the self-consistent NEGF-Poisson solver, $\Sigma^C_j$ is the self-energy of the $j^{th}$ contact (where $j\in R$ (Right), $L$ (Left)), which are calculated from the corresponding broadening functions ($\Gamma_{1,2}$), and $\Sigma^S$ is the scattering self-energy evaluated using the self-consistent Born approach \cite{Henrickson,Aeberhard_jce,QCLnegf,AeberhardPRB2008,PRL_nanotube,myTED,myPRA,mukherjee2018improved,priyadarshi2018superlattice,mukherjee2021carrier}.
We calculate the LDOS as the diagonal elements of the spectral function, given by \cite{mukherjee2021carrier,DattaQT}
\begin{equation}
	\mathcal{A}(z,z',E)=i \left[ G(z,z',E)-G^{\dagger}(z,z',E) \right],
	\label{LDOS_eq}
\end{equation}
for the entire energy ($E$) range of interest. In the ballistic transport limit, the electronic transmission probability is calculated as \cite{dattaLNE,DattaQT}
\begin{equation}
	T(E)=Re[Tr(\Gamma_{1}G\Gamma_{2}G^{\dagger})].
	\label{trans_eq}
\end{equation}
However, in practical devices, scattering effects caused due to the fluctuations at the atomic level, have a deep impact on the carrier transport. These effects destroy the system coherence and are included in the simulation setup through an additional self-energy coupled to the electron and hole correlation functions ($G^{n,p}$) which are given by
\begin{equation}
	\begin{split}
		&G^n(E)=G(E)\Sigma^{in}(E)G^{\dagger}(E),\\ &G^p(E)=G(E)\Sigma^{out}(E)G^{\dagger}(E),
	\end{split}
	\label{GnGp}
\end{equation}
where $\Sigma^{in}(E)=\sum_{j=L,R}\Gamma^C_j(E)f_j(E-\mu_j)+\Sigma^{in}_S(E)$ and $\Sigma^{out}(E)=\sum_{j=L,R}\Gamma^C_j(E)\left(1-f_j(E-\mu_j)\right)+\Sigma^{out}_S(E)$. In particular, $f_j(E-\mu_j)$ is the equilibrium Fermi function corresponding to the $j^{th}$ contact with the electrochemical potential $\mu_j$ and $\Sigma_S^{in(out)}$ signifies the in (out) scattering functions related to the scattering event $S$. \\
\indent The scattering matrices corresponding to the momentum and phase relaxed elastic scattering events are considered as $\Sigma_S^{in(out)}=DG^{n(p)}$, where the scaling factor D signifies the strength of scattering which in our simulation is taken as $D= 10^{-5} eV^2$. In contrast, the high temperature energy relaxing inelastic scattering processes are associated with the absorption and emission of phonons. The scattering rates in such cases are determined from the first order self-consistent Born's approximation (SCBA), given by
\begin{equation}
\Sigma_S^{in}(E)=D_0(N_{w}+1)G^{n}(E+\hbar*w)+ D_0N_{w}G^{n}(E-\hbar*w),
\end{equation}
\begin{equation}
\Sigma_S^{out}(E)=D_0(N_{w}+1)G^{p}(E-\hbar*w)+D_0(N_{w})G^{p}(E+\hbar*w)
\end{equation} where, $N_{w}=(\exp(\frac{\hbar*w}{K_{B}T})-1)^{-1}$ represents equilibrium phonon occupation number at temperature T with frequency $w$. Here, the inelastic scattering coefficient $D_{0}$ is taken as $10^{-2}$ $eV^2$.
In such cases, \eqref{trans_eq} no longer remains valid and the spectral current flowing from point $z_{i}$ to $z_{i+1}$ is then given by
\begin{equation}
	I^{sp}_{e(h)}(E)=\frac{iq}{\pi \hbar}\left[ H_{i,i+1}G^{n(p)}_{i+1,i}(E)\\-H_{i+1,i}G^{n(p)}_{i,i+1}(E) \right].
	\label{Ieqn}
\end{equation}
A detailed description of the carrier scattering and their modeling in the NEGF framework can be found in Refs. \cite{scattering_negf,nikonov2009scattering,DattaQT}. In our study, we consider both the elastic and inelastic models of scattering based on their importance and validity.

\section{RESULTS AND DISCUSSIONS}
\label{sec_result}
\subsection{Electronic band properties}
\label{eg_kp} 
\begin{figure}[!htbp]
	\centering
	{\includegraphics[height=0.2\textwidth,width=0.45\textwidth]{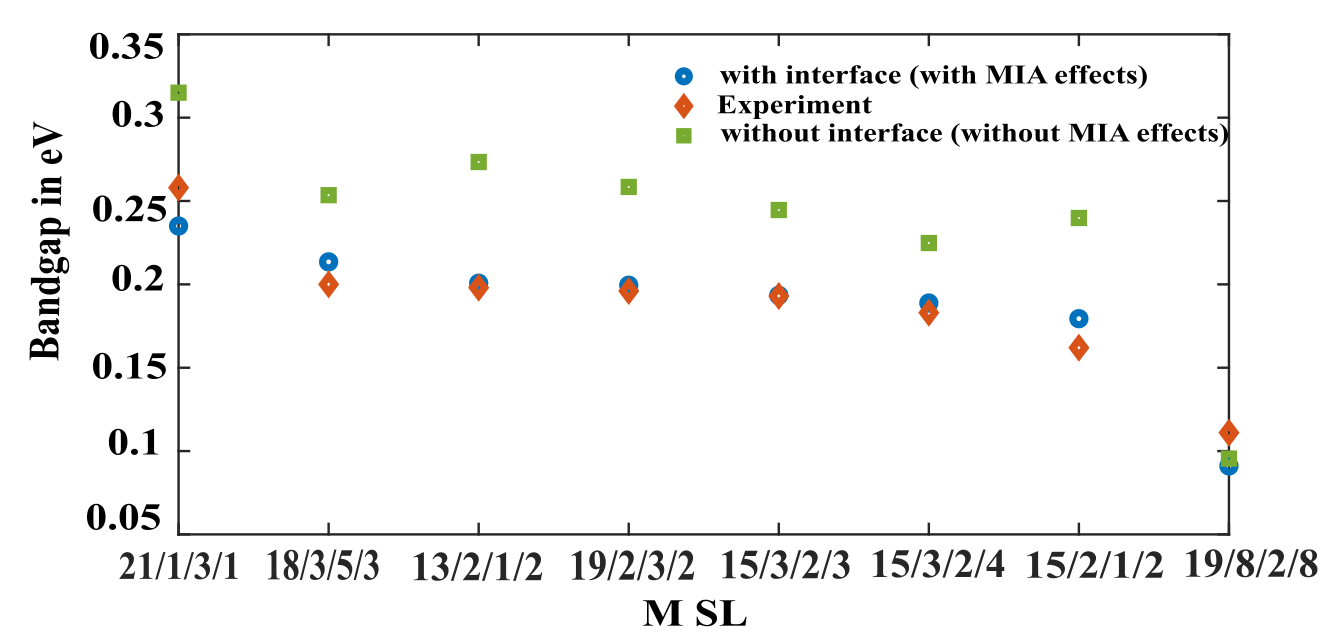}\label{E3}}
	\caption{Comparison between measured and theoretically predicted bandgaps of eight MSL samples. Without MIA effects the bandgaps were largely overestimated, whereas with the inclusion of MIA effects, they are obtained in close proximity to the experimental values. }\label{Bg}
\end{figure}

\begin{figure}[!htbp]
	\centering
	\subfigure[]{\includegraphics[height=0.225\textwidth,width=0.225\textwidth]{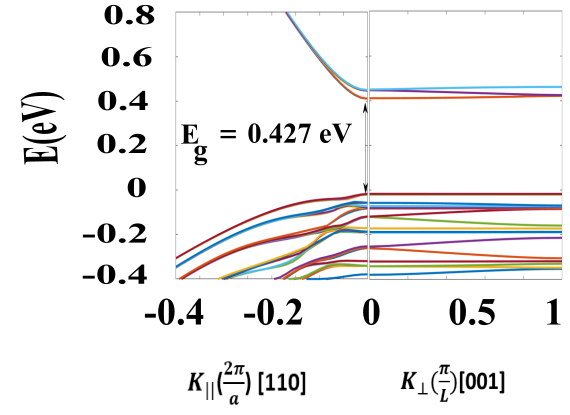}\label{E1}}
	\quad
	\subfigure[]{\includegraphics[height=0.225\textwidth,width=0.225\textwidth]{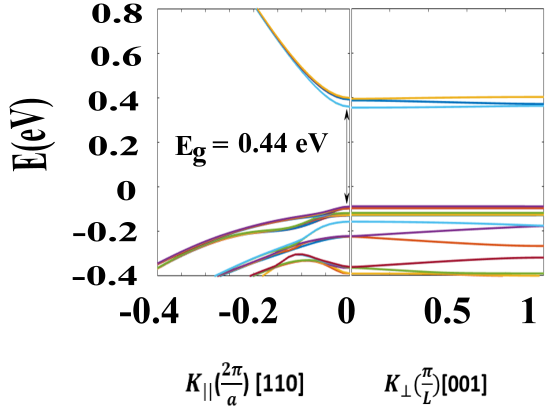}\label{E2}}

	\caption{Electronic band structure of T2SL and  MSL at T = 77 K using 8  band $k.p$ method: In-plane and out-of-plane E-k dispersion within the 1st Brillouin zone are calculated for (a) 6ML/20ML T2SL and (b) 7ML/6ML/3ML/6ML MSL using the periodic boundary condition with $H_{IF}$ matrix added at the interface. The obtained plots depict strong anisotropy between the in-plane and out-of-plane directions of the both the structures having almost similar bandgaps.}
	\label{Ekp}
\end{figure}

\begin{figure}[!htbp]
	\centering
	\subfigure[]{\includegraphics[height=0.25\textwidth,width=0.4\textwidth]{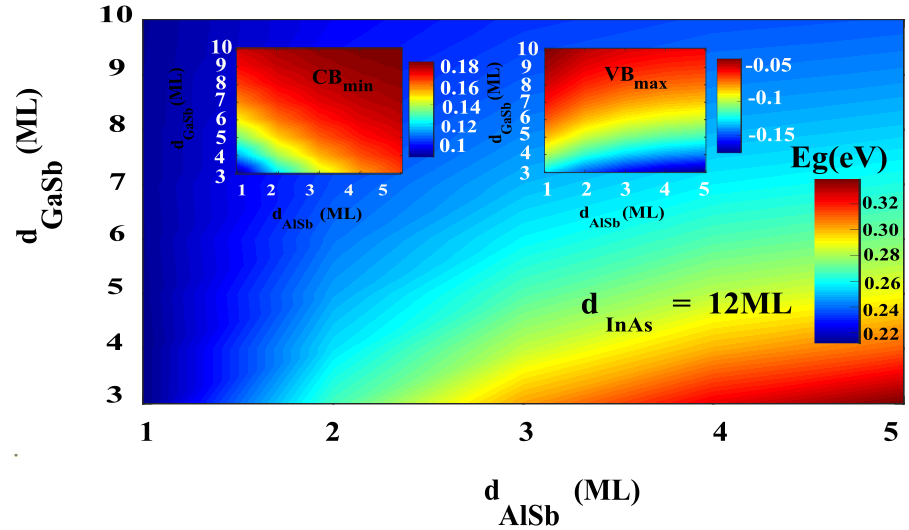}\label{bg1}}
	\quad
\quad
\subfigure[]{\includegraphics[height=0.25\textwidth,width=0.4\textwidth]{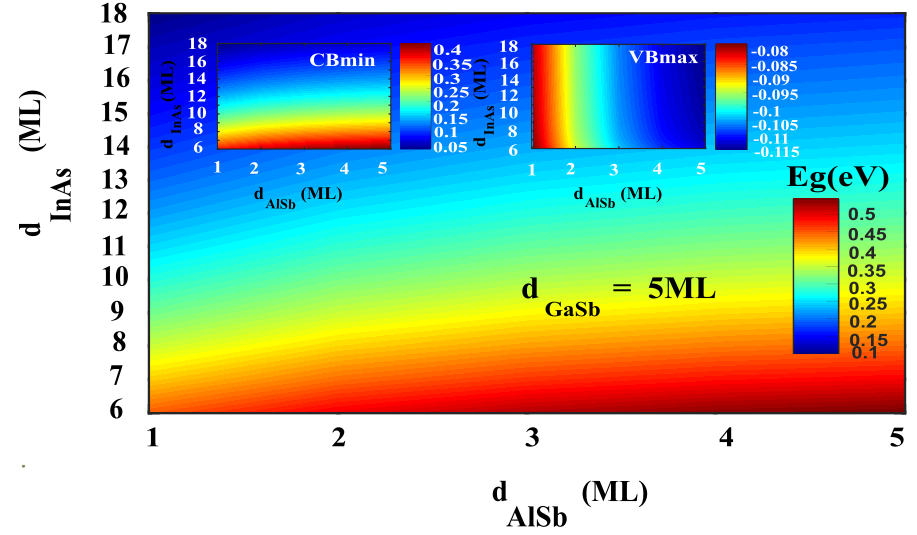}\label{bg2}}
	\caption {Bandgap and band offsets of MSL with reference to zero  energy level at $T=77K$ in 2D color plots: variation of $E_{g}$ with respect to widths of (a) $d_{AlSb}$ and $d_{GaSb}$  when $d_{InAs}$ is kept constant, and (b) $d_{AlSb}$ and $d_{InAs}$ at constant $d_{GaSb}$. In the former case, VB\textsubscript{max} and CB\textsubscript{min} shift in opposite directions with the increase in AlSb thickness, and CB\textsubscript{min} is less sensitive with the change in AlSb thickness unlike VB\textsubscript{max}. The latter case portrays the inverse effect of InAs width on CB\textsubscript{min}, while  AlSb layer width has direct impact on VB\textsubscript{max}, offering a superior tuning of band edge alignments.}\label{Eg}
\end{figure}

\begin{figure}[!htbp]
	\centering
	\subfigure[]{\includegraphics[height=0.2\textwidth,width=0.2\textwidth]{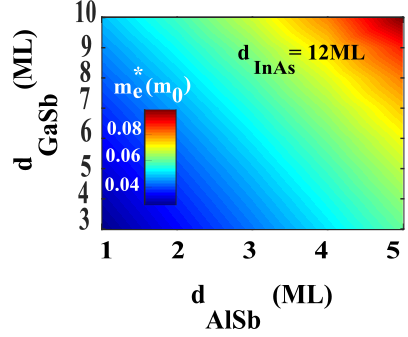}\label{Eg_m1}}
	\quad
	\subfigure[]{\includegraphics[height=0.2\textwidth,width=0.2\textwidth]{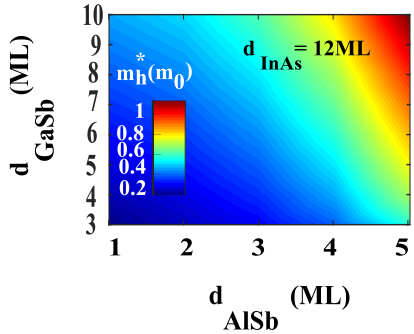}\label{Eg_m2}}
	\quad
		\subfigure[]{\includegraphics[height=0.2\textwidth,width=0.2\textwidth]{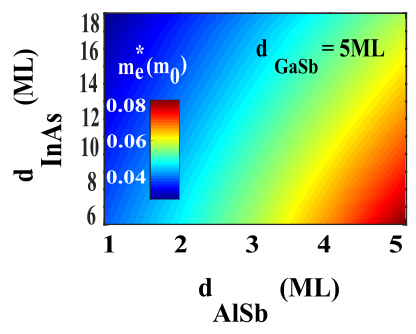}\label{Eg_m3}}
	\quad
		\subfigure[]{\includegraphics[height=0.2\textwidth,width=0.2\textwidth]{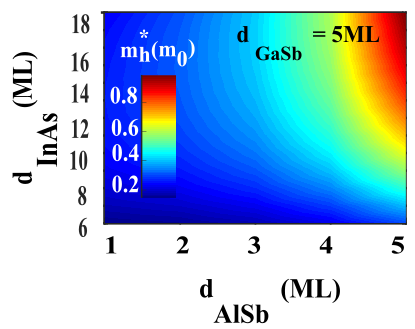}\label{Eg_m4}}
		\quad
	\caption{MSL electron and hole effective masses with 2D color plots at $T=77K$. (a) and (b) Varying $d_{GaSb}$ and $d_{AlSb}$ while keeping $d_{InAs}$ as constant. (c) and (d) Varying $d_{InAs}$ and $d_{AlSb}$ while keeping GaSb as constant. AlSb barrier makes electron wavefunction more localised in InAs wells, therefore $m^*_e$ increases. $m^*_h$ increases marginally with AlSb at lower widths of InAs and GaSb, but rises sharply at higher widths of InAs and GaSb. $m^*_e$ and $m^*_h$ increase with AlSb thickness,  therefore the tunnelling probability of carriers will be less in these MSL configurations.}
	\label{m_1}
	
\end{figure}

The peak absorption properties and the spectral range of operation are the most crucial features of any IR detector setup which can be determined from the knowledge of the band structure of the absorber material. In particular, the bandgap, bandwidth and the other key parameters derived from the band structure hold much significance owing to their direct connection with the aforementioned detector parameters.\\ \indent We, therefore, start the discussion of this section by first evaluating the band structure properties of a MSL structure and compare the same with that of a T2SL having almost similar bandgap and electron effective mass for the sake of a comprehensive understanding about their key differences \cite{lang2013electronic}.
In order to achieve the correct bandgap values in these broken band alignment structures having interfaces formed by NCA, the appropriate interface treatment is required to account for the MIA  effects \cite{li2010intrinsic}.
For this, we have considered the interface matrix at the junction  of InAs and GaSb layers, as discussed in section \ref {sec_1}. In order to validate the k.p model with interface consideration for MSL, we calculate the band gap for some MSL configurations, and found that the obtained bandgaps with the MIA effects are in close agreement with the experimental values as shown in Fig. \ref{Bg}. Next, employing the same model, we calculate the electronic band structures of 6ML/20ML T2SL and 7ML/6ML/3ML/6ML MSL structures having almost similar bandgaps and plot the obtained $E-k$ dispersion results at T=77K in Figs. \ref{E1} and \ref{E2}. We consider a supercell of three periods to capture the impact of the alternative interface potentials \cite{li2010intrinsic}. Here, due to the double degeneracy \cite{becer2019modeling}, each state splits into two states. The obtained effective masses from the curvature of dispersion plots for T2SL are $m^*_e=0.0466m_{0}$ and $m_h^*=0.258m_{0}$, similarly for MSL, $m_e^*=0.0588m_{0}$ and $m_h^*=0.26m_{0}$.\\
\indent In T2SL, the electrons (holes) are spatially confined in InAs (GaSb) layers and their eigen energies vary with the width of that layer \cite{lang2013electronic,ting2020long,delmas2019comprehensive}. It is evident that the change in conduction band energy level is eminent and sensitive with respect to the thickness of the InAs layer, leading to a robust control on the conduction band offset \cite{lang2013electronic,ting2020long}. However, the variation of GaSb layer width does not have adequate control on the tuning of valence band due to the large heavy hole effective mass \cite{lang2013electronic,nguyen2008band,haugan2004band,mukherjee2021carrier}. Therefore, the T2SL system strongly suffers from valence band tunability which is overcome in MSL structures by inserting a thin AlSb layer within the GaSb layer \cite{lang2013electronic,ting2020long}.  \\
 \indent The insertion of AlSb divides the GaSb hole quantum well into two quantum wells, which leads to a reduction of the individual well widths. This suggests that both the GaSb and AlSb layers have a major role to play in tuning the valence band maxima (VB\textsubscript{max}) and thereby tailoring the bandgap. Figure. \ref{bg1} depicts the variation of bandgap  ($E_g$) with respect to $d_{GaSb}$ and $d_{AlSb}$ in a 2D color plot with a constant $d_{InAs}=12$ML . Similarly, Fig. \ref{bg2} shows the $E_g$ variation with  $d_{InAs}$ and $d_{AlSb}$ with a constant $d_{GaSb}=5$ML. The variation of conduction band minima (CB\textsubscript{min}) and VB\textsubscript{max} are also shown with respect to the same parameters in the insets of Fig. \ref{bg1} and \ref{bg2}.
 The presence of the additional AlSb barrier reduces the interaction between the electrons confined in adjacent InAs layers and as a consequence CB\textsubscript{min} moves upwards due to the lowering of conduction band split \cite{haugan2004band} as evidenced from Fig. \ref{bg1}.
 Moreover, the electron wavefunction becomes more localized in InAs wells which effectively gives rise to a higher electron effective mass in MSL \cite{lang2013electronic,nguyen2008band,haugan2004band} as shown in Fig. \ref{m_1}.
 In Fig. \ref{bg1}, VB\textsubscript{max} shifts up (down) with the increase in $d_{GaSb}$ ($d_{AlSb}$) when  $d_{AlSb}$ ($d_{GaSb}$) is kept constant. Furthermore, it is observed that as the AlSb width goes up, the lowest attainable limit of VB\textsubscript{max} increases even when GaSb thickness remains constant. For instance, at $d_{AlSb}=5ML$ and GaSb varying from 3ML to 10ML, the lowest value attained by VB\textsubscript{max} is found to be around $-0.2 eV$, whereas at $d_{AlSb}=1ML$ this value reaches only $-0.125 eV$.
 $d_{InAs}$ has more pronounced effect on CB\textsubscript{min} than $d_{AlSb}$, therefore, CB\textsubscript{min} shown in the inset of Fig. \ref{bg2}, shifts downwards with the increase in $d_{InAs}$, and there is almost no change with $d_{AlSb}$.
It appears in Fig. \ref{bg2} that at a constant $d_{AlSb}$ and increasing $d_{InAs}$ from 6ML to 18ML, VB\textsubscript{max} remains unchanged, which is similar to the trend noticed in T2SL \cite{delmas2019comprehensive,mukherjee2021carrier}. Here, VB\textsubscript{max} moves from -0.08 $eV$ to -0.115$eV$ as $d_{AlSb}$ is increased from 1ML to 5ML for all values of $d_{InAs}$ under consideration. This suggests that VB\textsubscript{max} changes sharply with both $d_{AlSb}$ and $d_{GaSb}$, however, remains nearly invariant with $d_{InAs}$. Furthermore, at a constant $d_{AlSb}$ and with increasing $d_{InAs}$, as  CB\textsubscript{min} is pulled down with VB\textsubscript{max} remains unchanged, the effective bandgap decreases as observed in Fig. \ref{bg2}. Such a wide tunable range in valence band edge is key to design hole blocking barriers in hetero-structure based photodetectors to suppress the SRH processes in the depletion region which in turn alleviates dark current \cite{nguyen2007dark,zavala2020antimonide,rodriguez2007n}. The obtained bandgap values in Fig. \ref{bg1} and Fig. \ref{bg2} corresponds to the wavelength range between 3$\mu m$ to 12$\mu m$.\\
\indent Next, we turn our attention towards the DOS effective masses of electrons and holes which are calculated from the band structure data using the relation $m_{e(h)}^*=\left( m_{e(h)\parallel}^*\right)^{2/3} \left( m_{e(h)\perp}^*\right)^{1/3}$ and are plotted in 2D color plots with respect to the variation in the layer thicknesses. It is seen from Fig. \ref{Eg_m1} and Fig. \ref{Eg_m3} that the obtained electron effective masses of MSL are higher than that of T2SL for a similar wavelength range \cite{mukherjee2021carrier,delmas2019comprehensive} due to the strong electron localization in InAs layer is fuelled by the additional AlSb layer \cite{lang2013electronic,nguyen2008band,haugan2004band}. In particular, when we consider the bandgap range from $0.1 eV$ to $0.3 eV$, the maximum electron mass obtained in case of T2SL is approximately $0.034m{0}$ \cite{mukherjee2021carrier,delmas2019comprehensive}, while in MSL, for the same range, the electron masses are varying from $0.06m_{0}$ to $0.08m_{0}$. It is also evident that higher electron effective masses can be achieved even with thin GaSb layer, as they are more sensitive to $d_{AlSb}$   \cite{lang2013electronic,haugan2004band}. From Fig. \ref{Eg_m2} and Fig. \ref{Eg_m4}, it is observed that the rise in hole effective masses with the increase in AlSb width is prominent only at higher GaSb and InAs thicknesses.  The acquired higher electron masses are important for the p-$\pi$-M-n type structures, in which, an additional MSL is inserted between the $\pi$ and n region of traditional p-$\pi$-n structure to restrict the carrier transport due to diffusion and tunneling at depletion region to lessen the dark current \cite{nguyen2007dark,nguyen2007type,delmas2019comprehensive}.
	\subsection{Comparison of finite MSL and T2SL absorber}
\label{comp}
\begin{figure}[!htbp]
	\centering
	\quad
	
		\subfigure[]{\includegraphics[height=0.19\textwidth,width=0.20\textwidth]{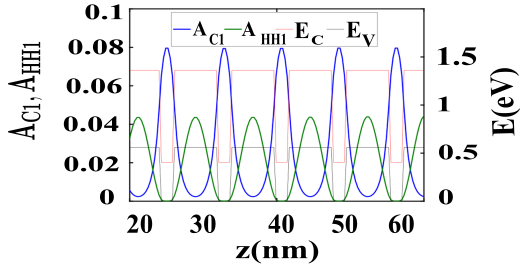}\label{s1}}
		\quad
		\subfigure[]{\includegraphics[height=0.19\textwidth,width=0.2\textwidth]{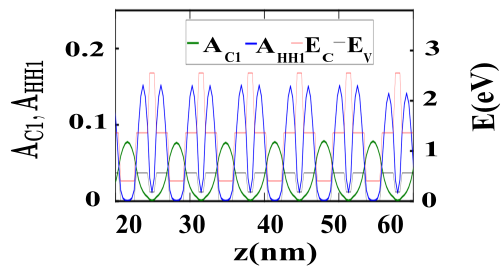}\label{s2}}
	\quad
		\subfigure[]{\includegraphics[height=0.2\textwidth,width=0.2\textwidth]{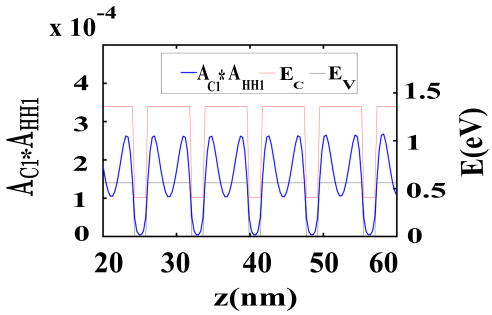}\label{s3}}
		\quad
			\subfigure[]{\includegraphics[height=0.2\textwidth,width=0.2\textwidth]{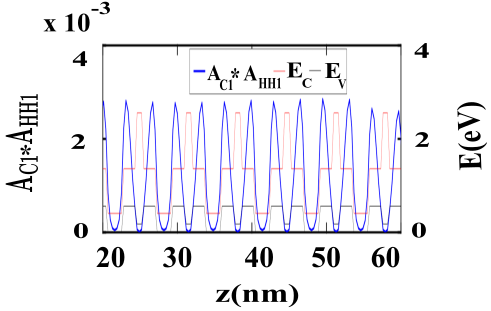}\label{s4}}
	\quad
\caption{ Available states in conduction and valence band and their spatial product at $T=77K$. (a) and (b) number of available states in conduction band and valence band with respect to the position for T2SL and MSL. (c) and  (d) The spatial product of number of available states in C1 and  HH1 for T2SL and MSL. For MSL, insertion of a thin AlSb layer in the middle of each GaSb hole quantum well pushes the HH1 wave function out for stronger overlap with the C1 wave function, therefore the interband overlap is higher in MSL, which indicates more absorption.}
	\label{states}	
\end{figure}

\begin{figure}[!htbp]
	\centering 
		\subfigure[]{\includegraphics[height=0.2\textwidth,width=0.2\textwidth]{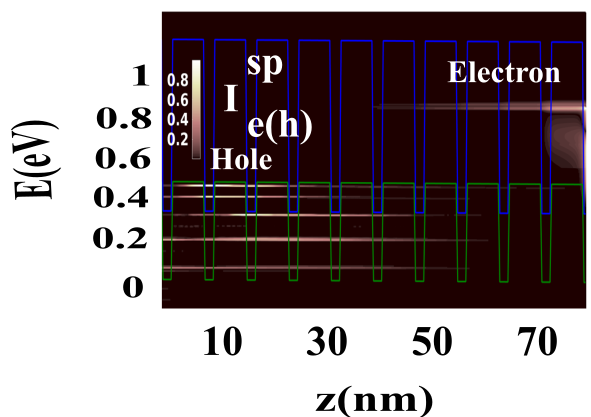}\label{c1}}
	\quad
	\subfigure[]{\includegraphics[height=0.2\textwidth,width=0.2\textwidth]{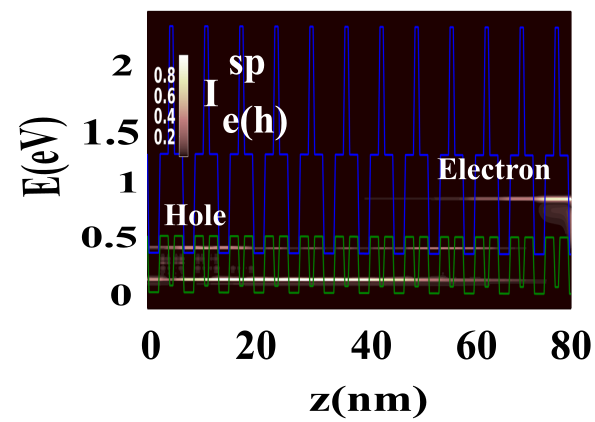}\label{c2}}
	\quad
	\caption{Spatially and energetically resolved normalized dark current for ten periods of 6ML/20ML T2SL and twelve periods of 7ML/6ML/3ML/6ML MSL with the inclusion of optical phonons with energy $30meV$ at $T=300K$ with the applied voltage of 0.015 V (a) Dark current spectrum in conduction and valence bands of T2SL (b) Dark current spectrum in conduction and valence bands of MSL. The dark current spectrum at lead-device interfaces reflects the spectrum of states from which the carriers injects from the contacts.
}	\label{cur}
\end{figure}

Our next step is to optimize the absorber layer configuration for better optical properties through the enhancement of interband carrier overlap \cite{ting2020long}.  Due to the spatial separation of carriers, T2SL has less oscillator strength, which leads to  weaker absorption \cite{ting2020long,razeghi2010band}. Whereas, the optical properties of  MSL are expected to be better than T2SL, and even comparable to short period T2SL \cite{taalat2013influence,lang2013electronic}. To model finite superlattice structures in the quantum mechanical framework, we implement single band NEGF approach using the effective mass Hamiltonian and on the basis of obtained properties, we compare the properties of few-period MSL and T2SL structures \cite{mukherjee2021carrier,priyadarshi2018superlattice,akhavan_effectivemass,akhavan2016superlattice,aeberhard2018photocarrier}.
The parameters used for the simulation are provided in Table \ref{negftable}.
\begin{table}
	\centering
	\caption{ Material parameters used in the NEGF simulation \cite{livneh2012k,vurgaftman2001band}}
	\begin{tabular}{|l|l|l|l|}
		\hline
		Parameters & InAs & GaSb & AlSb \\
		\hline
		Electron effective mass ($m_e^*$) & 0.023 & 0.041 & 0.14 \\
		\hline
		Heavy hole effective mass ($m_h^*$) & 0.4 & 0.4 & 0.9\\
		\hline
		VBO [$eV$] @300K & -0.50 & 0 & -0.44\\
			VBO [$eV$] @77K & -0.56 & 0 & -0.38\\
		\hline
	\end{tabular}
	\label{negftable}
\end{table}
 \begin{figure}[!htbp]
	\centering

	\subfigure[]{\includegraphics[height=0.2\textwidth,width=0.2\textwidth]{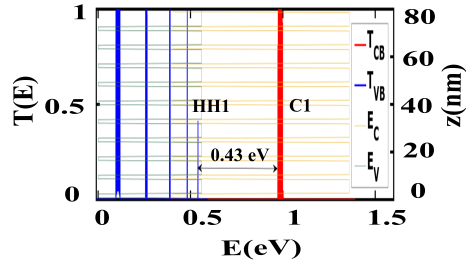}\label{N1}}
	\quad
	\subfigure[]{\includegraphics[height=0.2\textwidth,width=0.2\textwidth]{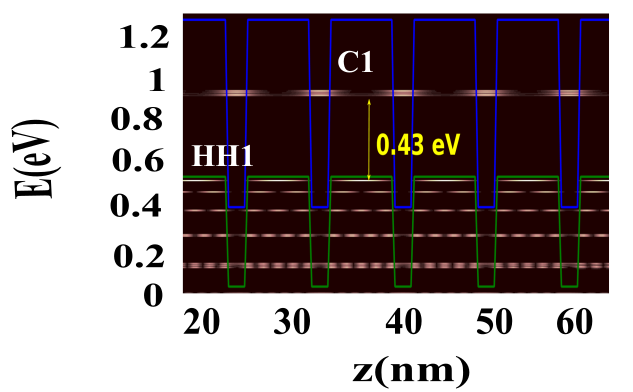}\label{N2}}
		
	\quad
\subfigure[]{\includegraphics[height=0.2\textwidth,width=0.2\textwidth]{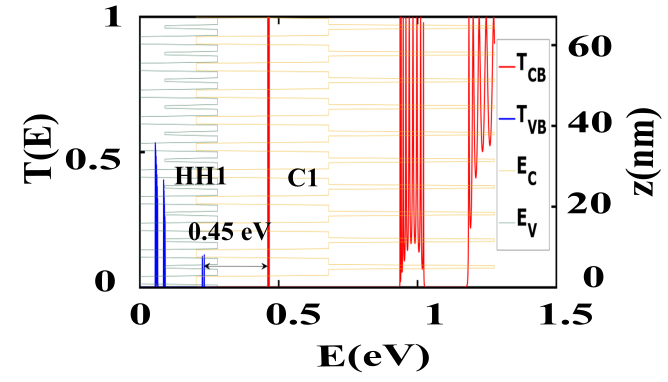}\label{N3}}
	\quad
	\subfigure[]{\includegraphics[height=0.2\textwidth,width=0.2\textwidth]{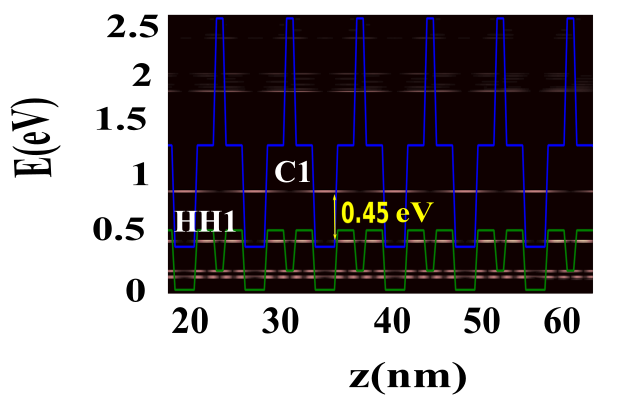}\label{N4}}
	\quad
	\caption{Transmission function and miniband formation at $T=77K$: (a) Electron and hole transmission probabilities of ten period 6ML/20ML T2SL. (b)  LDOS of ten period 6ML/20ML T2SL. (c) Electron and hole transmission probabilities of a 10 period 7ML/6ML/3ML/6ML MSL. (d) LDOS of ten period 7ML/6ML/3ML/6ML MSL. Transmission probabilities are plotted with respect to energy, and  LDOS plotted in a grey scale 2-D plot in the position and energy space. The transmission probability in MSL is less than T2SL in valence band. The conduction band bandwidth of MSL is less than T2SL, also there is a  formation of  two quantum wells in GaSb due to the AlSb layer in MSL.}
		\label{NEGF1}
\end{figure} 
Figure \ref{N1} and \ref{N3} depict the electron and heavy hole transmission probabilities in the ballistic limit for the same configuration of T2SL and MSL used in k.p calculations, respectively \cite{priyadarshi2018superlattice,mukherjee2018improved}.
To obtain absolute transmission, the conduction and valence band edges of the contacts are assumed at their respective lowest and highest values. In Fig. \ref{N2} and Fig. \ref{N4}, LDOS obtained via \eqref{LDOS_eq} within the elastic limit of scattering are shown for both T2SL and MSL \cite{priyadarshi2018superlattice,mukherjee2021carrier}.
The electron (hole) localization in InAs (GaSb) layer is fairly evident from the LDOS plots. In particular, the hole confinement in the two adjacent GaSb quantum wells formed due to the insertion of AlSb, can be distinctly seen in MSL. In addition, the first conduction mini bandwidth observed in MSL is lower than that of T2SL, which implies for higher localization of electrons in MSL. The bandgaps obtained through the NEGF simulation for the two structures are similar and in well agreement with the $k.p$ results, discussed in Sec. \ref{eg_kp}.\\
\indent In  Fig. \ref{s1} and Fig. \ref{s2}, we plot the total number of states in the first conduction ($A_{C1}$) and heavy hole ($A_{HH1}$) bands for T2SL and MSL, respectively, as a function of position by integrating the LDOS over the energy range of interest \cite{mukherjee2021carrier,priyadarshi2018superlattice}. In continuation, we further plot the spatial product ($A_{C1}$*$A_{HH1}$) of them in Fig. \ref{s3} and Fig. \ref{s4} to understand the nature of interband overlap responsible for carrier transition \cite{mukherjee2021carrier}.
 \begin{figure}[!htbp]
 {\includegraphics[height=0.2\textwidth,width=0.45\textwidth]{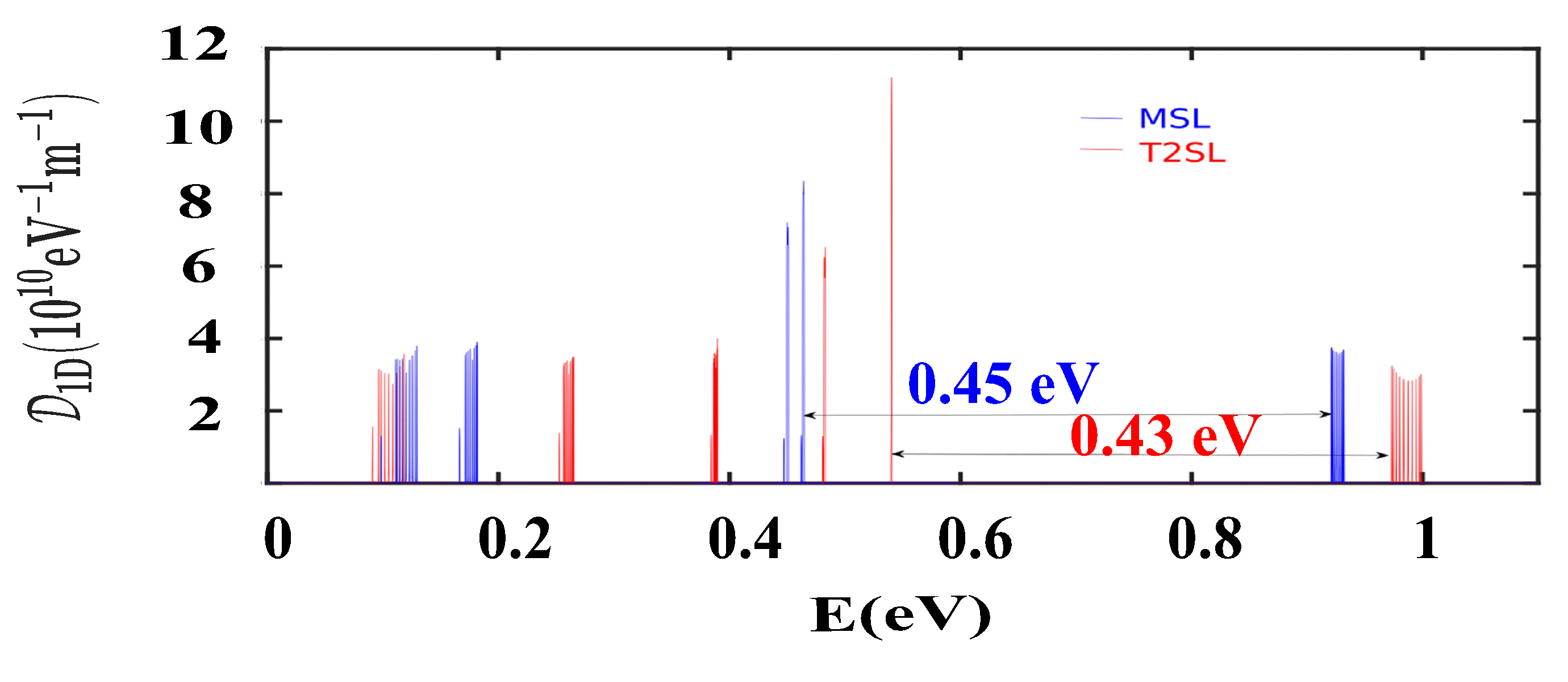}
 \caption{ 1D-DOS for first conduction and first heavy hole band for 6ML/20ML T2SL (red) and 7ML/6ML/3ML/6ML MSL (blue) at $T=77K$. The higher hole densities in T2SL suggests higher probability of auger recombination than MSL.}\label{l1}}
\end{figure}
This spatial product in MSL is found to be quite higher than that in T2SL especially at the interface region, which clearly indicates a stronger overlap between C1 and HH1 wavefunctions. The AlSb layer in MSL pushes the HH1 wavefunction towards the newly created wells. As a consequence, the centre of HH1 wavefunction is shifted more towards the center of C1 wavefunction which leads to an enhancement in the interband overlap, and hence provides higher absorption in MSL \cite{ting2020long}. In Fig. \ref{l1}, we plot the 1D-DOS calculated by integrating the LDOS over the entire T2SL and MSL lengths \cite{mukherjee2021carrier}. It is noticed that the 1D-DOS for VB holes are higher in T2SL than in MSL. Therefore,  by comparing the maximum hole density values in T2SL and MSL, we predict that Auger recombination will be less in MSL, causing lesser dark current \cite{flatte1998auger}. Also, higher electron localization in MSL is an indicative for reduced tunneling dark current in MSL absorber.\\
\indent Having evaluated and compared the LDOS properties of MSL and T2SL structures, it is now customary to examine the nature of carrier transport through these structures in order to gain a qualitative insight on the dark current at any given operating point and establish its connection to the LDOS. In doing so, we plot the normalized spectral current, given by \eqref{Ieqn}, of electrons and holes with respect to the position at an applied voltage of $0.015V$ and $T=300K$ in Fig. \ref{c1} and Fig. \ref{c2} for the T2SL and MSL structures, respectively. For a fair comparison, we maintain a nearly similar electric field corresponding to the applied bias of $0.015V$ across both these structures by considering ten periods of 6ML/20ML T2SL and twelve periods of 7ML/6ML/3ML/6ML MSL. The contacts are assumed to be carrier selective with one for the electron injection and the other for the hole. We also bring in the non-coherent transport features in our simulation model by including the self-consistent inelastic electron-phonon scattering model \cite{scattering_negf} with a phonon energy of $30meV$ \cite{deacon2005high,li2010intrinsic} to look for possible broadening of the current spectra, although they are less likely in the near-equilibrium regime of transport considered here. The bright stripes observed in both the figures are distinctly indicative of the miniband transport of electrons and holes for both the T2SL and MSL. However, a careful observation reveals that the T2SL current spectrum is broader than the MSL, especially near the contacts, which fairly justifies the strong localization in MSL as discussed earlier. Furthermore, the amount of broadening observed around the contacts with inelastic scattering largely vanishes when only the ballistic and elastic scattering effects are considered. This clearly points towards the existence of phonon-actuated energy relaxation processes of carriers occurring at higher temperatures. These processes become predominant at higher built-in and applied field and play a key role in tailoring the carrier transport. Therefore, this study holds much significance in the context of understanding the miniband and phonon-mediated hopping transport of carriers through heterostructures and can be extremely useful to predict the dark and photo current of T2SL systems. 

\subsection{MSL as unipolar or bipolar barrier}
\label{Barrier}
\begin{figure}[!htbp]
	\centering

	\quad
	\subfigure[]{\includegraphics[height=0.2\textwidth,width=0.2\textwidth]{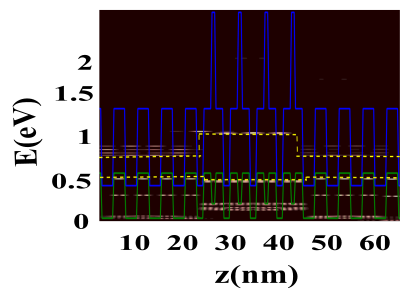}\label{b1}}
	\quad
				\subfigure[]{\includegraphics[height=0.2\textwidth,width=0.2\textwidth]{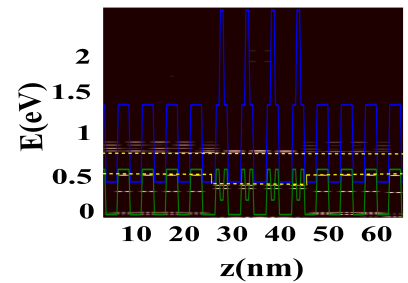}\label{b2}}

		\subfigure[]{\includegraphics[height=0.2\textwidth,width=0.2\textwidth]{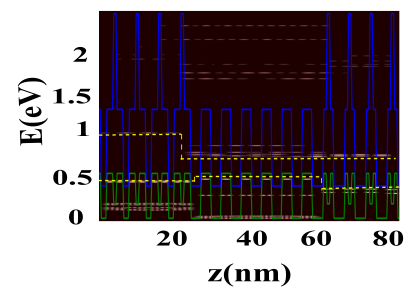}	\label{b3}}

		\caption{CB and VB offset tuning using MSL in 2D color LDOS plot at $T=77K$: unipolar (a) electron barrier ($B_e$) having CB offset of 0.27$eV$ using 5ML/6ML/3ML/6ML MSL and (b) hole barrier ($B_h$) with 0.13$eV$ VB offset using 10ML/3ML/3ML/3ML MSL are shown with respect to a 9ML/9ML T2SL absorber ($A_b$). (c) Bipolar barrier using $B_{e}A_bB_{h}$ design having both electron and hole blocking barriers. Specific band offsets can be attained by the appropriate tuning of thicknesses of multiple layers.
		}
	\label{bar}
	
\end{figure}
Photodetectors suffer from noise-inducing currents such as generation currents related to SRH processes in the depletion region, current due to thermal generation of carriers in the absorber region and their diffusion to the contact layers, and surface currents  \cite{maimon2006n,rakovska2000room}. SRH current, which dominates at lower temperature through the activated midgap traps, is minimized by utilizing barrier-based structures like nBn, XBp, XBn etc. The barrier blocks the majority carriers flow to inhibit the SRH processes in the depletion region \cite{zavala2020antimonide,rodriguez2007n}.
These barriers are usually made up of bulk materials which have limited applications due to their inadequacy to provide specific conduction and valence band tunability, and large dark currents at elevated temperatures \cite{zavala2020antimonide,ting2020long}. \\
\indent The design of barriers in the photodetector is noteworthy as the barrier height and width mutually determine the ability to block the thermal excitation of majority carriers from the contact layers, and the potential to impede the electron tunneling through it \cite{maimon2006n,zavala2020antimonide,nguyen2007type,nguyen2007dark,nguyen2009minority}. Here, we construct MSL based barriers as the insertion of AlSb layer provides an additional degree of freedom to design and control the band offsets as per requirement. Moreover, MSL structures due to their high effective mass as discussed in Sec. \ref{eg_kp}, become more resistant to the diffusion and tunneling transport in the depletion region \cite{nguyen2009minority,nguyen2007dark,nguyen2007type,lang2013electronic}.\\
\indent In this work, we present three configurations for $XB_{e}A_{b}$, $XB_{h}A_{b}$ and $XB_{e}A_{b}B_{h}X$ structures, where $B_{e}$ ($B_{h}$) is the electron (hole) barrier layer made of MSL, $A_{b}$ is the T2SL absorber layer and $X$ is the contact layer which can be composed of T2SL or bulk material. The band offsets pertaining to the electron and hole barriers with respect to a 9ML/9ML T2SL absorber ($A_b$), as obtained from the outcome of LDOS calculated via NEGF at 77$K$, are respectively plotted in Fig. \ref{b1} and Fig. \ref{b2}. Here, $B_e$ and $B_h$ are respectively modeled using 5ML/6ML/3ML/6ML and 10ML/3ML/3ML/3ML MSL configurations. It is noted from Fig. \ref{b1} that the $B_e$ layer provides a conduction band offset of approximately 0.27 $eV$ above and a nearly zero valence band discontinuity with respect to $A_{b}$, making it an ideal unipolar electron barrier that blocks the majority carriers electrons from the contacts and allows the minority holes to pass through it, functioning similar to a pn junction space charge region \cite{zavala2020antimonide,maimon2006n}. These barriers are usually intrinsic or have the similar doping as in the absorber region. Therefore, most of the depletion region lies within them, resulting in a reduced SRH dark current due to their high bandgap \cite{zavala2020antimonide,maimon2006n}. Similarly, $B_h$ in Fig. \ref{b2} offers a valence band offset of 0.13$eV$ with respect to $A_b$, while having a negligible conduction band discontinuity. Such a unipolar hole barrier opposes the flow of majority holes without affecting the minority electrons flow. Combining these two, one can design a $B_{e}A_{b}B_{h}$ bipolar barrier structure as depicted in Fig. \ref{b3} where $B_{e}$ ($B_{h}$) is sandwiched between the p-type (n-type) contact and absorber layer, which blocks the minority diffusion electron (hole) current from the p-type (n-type) contact \cite{gautam2013band,zavala2020antimonide,maimon2006n}. \\
\indent The design strategy to achieve such MSL barriers although appears to be quite challenging from an engineering perspective, follows from a definite physics-based guideline demonstrated earlier while discussing the band structure in Sec. \ref{eg_kp}. With reference to the 9ML/9ML T2SL, the AlSb layer in MSL splits the GaSb hole quantum well into two and reduces their effective width \cite{lang2013electronic}. This pushes VB\textsubscript{max} of MSL down with respect to the T2SL, giving rise to a VB offset for the configuration 10ML/3ML/3ML/3ML shown in Fig.\ref{b2}. In this case, for a zero CB offset, the thickness of the InAs layer in MSL should be kept slightly larger (10ML) than in T2SL to compensate the rise of CB\textsubscript{min} caused due to the presence of the AlSb electron barrier layer. Similarly, the lowering of InAs thickness in MSL pushes the CB\textsubscript{min} up and gives rise to a CB offset for the configuration 5ML/6ML/3ML/6ML as shown in Fig. \ref{b1}. To ensure the alignment of VB in this case to have zero offset, one should increase the thickness of the two GaSb layers (6ML each) to pull the VB\textsubscript{max} up to compensate the downshift in  VB\textsubscript{max} due to AlSb. This discussion, as a whole, should serve as a predictive and robust guideline to design efficient barriers using complex T2SL structures.
\section{Conclusion}
\label{conclu}
This study provided comprehensive design guidelines of utilizing M-structured superlattices for both the absorber and barrier layers through proper band engineering and discussed its potential benefits over conventional T2SL structures. Our detailed calculations carefully took into account the effects of both strain and microscopic interface asymmetry to primarily estimate the bandgap and density-of-states effective mass and their variation with respect to the thicknesses of the constituent material layers. In contrast, for practical finite-period structures, the local density-of-states and spectral tunneling transmission and current calculated using the Keldysh non-equilibrium Green's function approach with the inclusion of non-coherent scattering processes offered deep insights into the qualitative aspects of miniband and localization engineering via structural variation. Our key results demonstrated how to achieve a wide infrared spectral range, reduce tunneling dark currents, induce strong interband wavefunction overlaps at the interfaces for adequate absorption, and excellent band-tunability to facilitate unipolar or bipolar current blocking barriers. This study, therefore, exemplifies the utilization of 6.1{\AA} material library to its full potential through the demonstration of band engineering in M-structured superlattices and sets up the right platform to possibly replace other complex superlattice systems for targeted applications.

\section*{Acknowledgments}
The authors acknowledge the funding from the PMRF PhD scheme of Ministry of Education, Government of India. This work is also supported by ISRO-IIT Bombay Space Technology Cell. The research and development work undertaken in the project under the Visvesvaraya Ph.D Scheme of the Ministry of Electronics and Information Technology (MEITY), Government of India, is implemented by Digital India Corporation (formerly Media Lab Asia).

\bibliography{reference}
\end{document}